\documentclass[a4paper]{jpconf}
\usepackage{graphicx}
\begin{document}
\title{Pasta Structures of Quark-Hadron Phase Transition in Proto-Neutron Stars}

\author{Nobutoshi Yasutake$^1$, Toshiki Maruyama$^2$, Toshitaka Tatsumi$^3$}{
 \address{$^1$ Division of Theoretical Astronomy, National Astronomical Observatory of Japan, 2-21-1 Osawa, Mitaka, Tokyo 181-8588, Japan\\
$^2$ Advanced Science Research Center, Japan Atomic Energy Agency, Tokai, Ibaraki 319-1195, Japan\\
$^3$Department of Physics, Kyoto University, Kyoto 606-8502, Japan}
}
\ead{yasutake@th.nao.ac.jp}

\begin{abstract}
We study the quark-hadron mixed phase in proto-neutron stars with the finite-size effects. 
In the calculations of pasta structures appeared in the mixed phase, the Gibbs conditions require the pressure balance and chemical equilibrium between two phases besides the thermal equilibrium.  
We find that the region of the mixed phase is limited due to thermal instability. 
Moreover, we study the effects of neutrinos to the pasta structures. 
As a result, we find that the existence of neutrinos make the pasta structures unstable, too. 
These characteristic features of the hadron-quark mixed phase should be important for the middle stage of the evolutions of proto-neutron stars. 
\end{abstract}

%------------------------
\section{Introduction}
%------------------------
There are many uncertainties left for the hadron-quark phase transition, e.g.\ the equation of state (EOS) of quark matter or deconfinement mechanism. 
Assuming the quark deconfinement transition to be of the first order, 
%a thermodynamical instability and 
%a mixed phase appear around the critical density. 
since there are two conserved quantities, baryon number and electric charge, the phase equilibrium in the mixed phase must be carefully treated 
by applying the Gibbs conditions \cite{glendenning92}, instead of the Maxwell construction. 
A simple treatment of the mixed phase may be the bulk Gibbs calculation, where phase equilibrium of two bulk matter is considered 
without electromagnetic interaction and surface tension.
Generally the properties of the mixed phase should strongly  depend on electromagnetic interaction and surface tension, 
and these  effects, sometimes called ``{\it the finite-size effects}", lead to the non-uniform
``Pasta" structures. 
The EOS of the mixed phase becomes similar to the one by the bulk Gibbs calculation for weak surface tension, 
and to the one given by the Maxwell construction for strong surface tension~\cite{voskresensky02,endo06,maruyama07,maruyama08}. 
The charge screening is also important for their mechanical instability. 
In this paper, we study the hadron-quark mixed phase with the finite-size effects at finite temperature by extending the previous works~\cite{maruyama07,maruyama08,yasutake09b}. We also present some results for the neutrino-trapping case and discuss its effect.

%This paper is organized as follows. In second section, we outline our framework%.
% In third section, we present numerical results. The last sections is the summa%ry and discussion. 

%-----------------------------------------------
\section{Equation of state for each phase}
%-----------------------------------------------
The theoretical framework for the hadronic phase of matter is the
nonrelativistic Brueckner-Hartree-Fock approach including hyperons such as
$\Sigma^-$ and $\Lambda$~\cite{baldo98}.
There is a controversy about the $\Sigma^-$-$N$ interaction. 
The recent experimental result about hypernuclei has suggested that it is repulsive~\cite{noumi02,saha04}, 
while we use here a weak but attractive interaction;  
$\Sigma^-$ then appears before $\Lambda$ in uniform and neutral matter. 
It would be interesting to see how our results are changed 
by using different $\Sigma^-$-$N$ interaction,
and we will discuss it in the future work.
Moreover, we adopt the frozen-correlation approximation at finite temperature~\cite{yasutake09b, baldo99}. 
This approximation is 
feasible for temperature $T < 50$ MeV.

For the quark phase, we adopt the MIT bag model or the density dependent bag model consisting of
$u,~d~,s$ quarks~\cite{nicotra06b}.
Probably, these models are too simple to describe quark matter in a realistic way.
We will adopt more sophisticated models in the future~\cite{yasutake09a}.
We assume massless $u$ and $d$ quarks and $s$ quarks with the current mass of $m_s=150$ MeV. 
We set the bag constant $B$ to be 100~MeV fm$^{-3}$ in the MIT bag model.
In the density dependent bag model, we set the vacuum energy $B(\rho)$ as 
\begin{eqnarray}
 B(\rho) = B_\infty +(B_0 - B_\infty) \exp[-\beta \frac{\rho}{\rho_0}] \nonumber
 \label{eq:01}
 \end{eqnarray}   
with $B_\infty = 50$ MeV fm$^{-3}$, $B_0=400$ MeV fm$^{-3}$, and $\beta=0.17$. 
For a more extensive discussion of this topic, the reader is referred to Nicotra et al.~\cite{nicotra06b} and references cited therein.

For the mixed phase, we assume non-uniform matter with various geometrical structures (``pastas"); droplet, rod, slab, tube, and bubble. 
Introducing the Wigner-Seitz cell to treat such non-uniform structure, we use the local density approximation for particles.
Between the hadron and quark phases we put a sharp boundary with  a constant surface tension parameter.
We impose the Gibbs conditions, i.e.\ chemical equilibrium among particles in two phases 
consistent with the Coulomb potential, 
and a pressure balance consistent with the surface tension.
Here, we write down the conditions of the  chemical equilibrium in the mixed phase;

\begin{eqnarray}
 && \mu_u+\mu_e - \mu_{\nu{_e}}= \mu_d = \mu_s , \nonumber\\
 && \mu_p+\mu_e - \mu_{\nu{_e}}= \mu_n = \mu_\Lambda = \mu_u + 2\mu_d ,\\
 && \mu_{\Sigma^-} + \mu_p = 2\mu_n .\nonumber
\label{eq:02}
\end{eqnarray}
Note that we do not take into account the anti-particles in this paper.

Since the role of the surface tension to the mixed phase has been already studied in the previous papers~\cite{maruyama07,maruyama08}, 
we use one constant value of the surface tension parameter $\sigma_S=$ 40 MeV~fm$^{-2}$ in this paper.

Finally, we compare the Helmholtz free energy among the pasta structures and choose which structure is most favored.  
The Helmholtz free energy for each cell is then given as
\begin{eqnarray}
F = \int_{V_H} dr^3 \mathcal{F}_H[n_i] 
 + \int_{V_Q} dr^3 \mathcal{F}_Q[n_q] + F_e + F_{\nu_e}+ E_C + E_S
\label{eq:03}
\end{eqnarray}
with $i=n,p,\lambda,\sigma$, $q=u,d,s$, 
$\mathcal{F}_H$~($\mathcal{F}_Q$) is the Helmholtz free energy density for hadron~(quark) matter, 
and $E_S=\sigma S$ the surface energy with $S$  being the hadron-quark interface area. 
$F_e$ and  $F_{\nu_e}$ are the free energies of the electrons and neutrinos.
For simplicity, muons are not included in this paper.

%------------------
\section{Results} 
%------------------
In proto-neutron stars, we must consider mainly two effects, namely temperature and neutrino fractions. The range of temperature is roughly $\sim$ 0 - 40 MeV, and the range of neutrino fraction is $\sim$ 0 - 0.2~\cite{burrows86}. In previous paper, we have studied the effects of temperature on the quark-hadron mixed phase~\cite{yasutake09b}. In this study, we check the effects of neutrino fractions and summarize the finite size effects of the quark-hadron mixed phase in the proto-neutron stars. 

%------------
\subsection{Effects of temperature}
\begin{figure}[h]
\includegraphics[width=18pc]{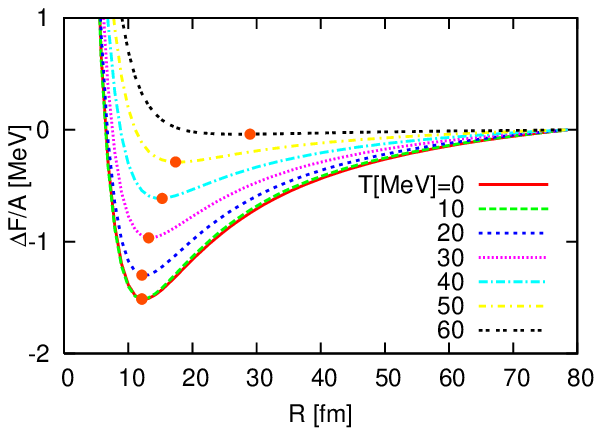}
\includegraphics[width=18pc]{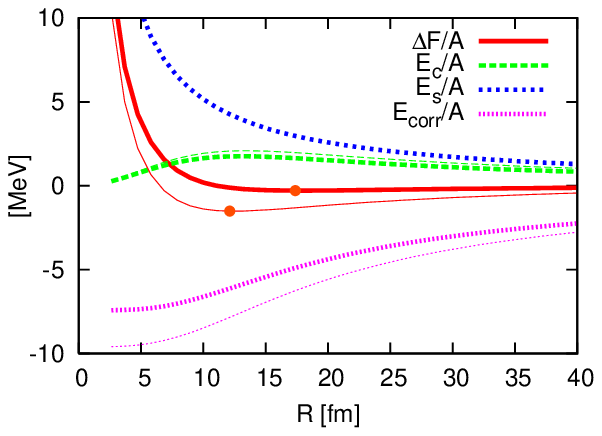}
\caption{\label{label01}The left panel shows that the droplet radius $R$ dependence of the free-energy per baryon for $n_B = 2~n_0$ and different temperatures. Here, we do not take into account neutrinos to see the effects of temperature. The quark volume fraction $(R/R_W)^3$ is fixed to be the optimal value at $T = 0$ MeV for each curve. The free energy is normalized by the value at $R \rightarrow \infty$. The filled circles on each curve shows the energy minimum. The results are for $B = 100$ MeV fm$^{-3}$, $\sigma= 40$ MeV fm$^{-2}$. The right panel shows each contribution to the $R$ dependence of the free energy, the Coulomb energy, the surface energy, or the correlation energy per baryon ($E_c/A$, $E_s/A$, $E_{corr}/A$) at $T = 0$ MeV (thick lines) and $T = 50$ MeV (thin lines). }
\end{figure}

In previous study, we have shown that finite temperature makes the mixed phase unstable~\cite{yasutake09b}.
Figure~\ref{label01} is one example to show the mechanical instability of the 
pasta structures as temperature is increased.
Here, we assume that the vacuum energy is constant, $B=100$ MeV fm$^{-3}$, applying the MIT bag model. 
The left panel of Fig.~\ref{label01} shows the free energies per baryon of the droplet structure for several values of temperature. The quark volume fraction in each cell $(R/R_W)^3$ is fixed to exclude the trivial $R$ dependence. Here we use the optimal value of $(R/R_W)^3$ at $T= 0$ MeV in every curve. 
We normalize them by subtracting the free-energy at infinite radius, $\Delta F = F(R)- F(R \rightarrow \infty)$, to show the $R$ dependence clearly. 
The structure of the mixed phase is mechanically stable below $T \sim 60$ MeV, but the optimal value of the radius $R$ is shifted to the larger value as $T$ increases. This behavior shows a signal of the mechanical instability and it comes from the charge screening effect and the thermal effect.
To elucidate this point more clearly, we show each contribution to $F/A$ in the right panel of Fig.~\ref{label01}; i.e., the Coulomb energy per baryon $E_C/A$, the surface energy per baryon $E_S/A$, and the correlation energy per baryon $E_{corr}/A$. The baryon density and the temperature are set as $n_B = 2~n_0$ and $T=50$ MeV. The correlation energy is produced by the difference of the bulk energy due to the rearrangement of charged particles. The entropy term in the bulk energy becomes dominant over the Coulomb energy in this case to exhibit the instability.
The details about the effects of temperature, including EOS,  are fully discussed in our previous paper~\cite{yasutake09b}.

%------------
\subsection{Effects of neutrinos}
\begin{figure}[h]
\includegraphics[width=18pc]{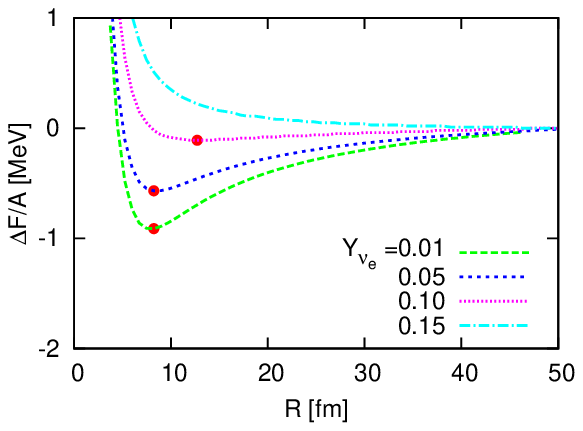}
\includegraphics[width=18pc]{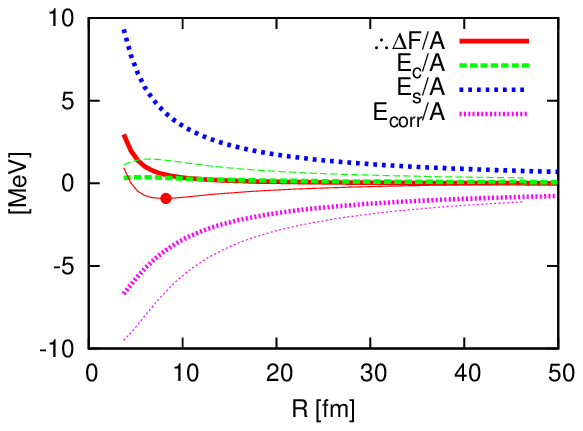}
\caption{\label{label02} Same as Fig.~\ref{label01}, but for different neutrino fractions $Y_{\nu_e}$ at constant temperature, $T=10$ MeV.  In this case, we set baryon density as $n_B = 2.5~n_0$, and the quark volume fraction $(R/R_W)^3$ is fixed to be the optimal value at $Y_{\nu_e} = 0.01$ for each curve. The thick lines of the right panel are the case of $Y_{\nu_e} = 0.15$, the thin lines $Y_{\nu_e} = 0.01$.}
\end{figure}

The left panel of Fig.~\ref{label02} shows the free energies per baryon of the droplet structure at several values of neutrino fractions. Here we use the optimal value of $(R/R_W)^3$ at $Y_{\nu_e}= 0$ in every curve. 
In these figures, we adopt the density dependent bag model shown as Eq.\ (\ref{eq:01}). 
We normalize them by subtracting the free-energy at infinite radius, as  same as Fig.~\ref{label01}. 
The structure of the mixed phase is mechanically stable below $Y_{\nu_e} \sim 0.10$, but the optimal value of the radius $R$ is shifted to the larger value as $Y_{\nu_e}$ increases. 

To elucidate this point more clearly, we check each contribution to $F/A$ in the right panel of Fig.~\ref{label02} again. The baryon density and the temperature are set to be $n_B = 2.5~n_0$ and $T=10$ MeV. In this, we compare the case for $Y_{\nu_e} = 0.01$~(thin lines) with the one for $Y_{\nu_e} = 0.15$~(thick lines). 
We can see the main contributions responsible to the change of free energy are the correlation energy and the Coulomb energy, though the thermal effect appears only in the correlation energy in Fig.~~\ref{label01}. 

Why does the Coulomb energy become important for the neutrino trapped case?
With the presence of neutrinos, the electron fractions becomes rich because of the chemical equilibrium Eq.\ (\ref{eq:02}).  Accordingly the fraction of particles with positive charge is enhanced through the charge neutrality condition. Thus, the net charge density is reduced over the whole region of the cell. Consequently  the Coulomb energy is largely reduced with neutrinos. The change of the correlation energy follows that of the Coulomb energy since the rearrangement effect is also reduced. Both the Coulomb and correlation energies give rise to the instability in this case.

\begin{figure}[h]
\includegraphics[width=20pc]{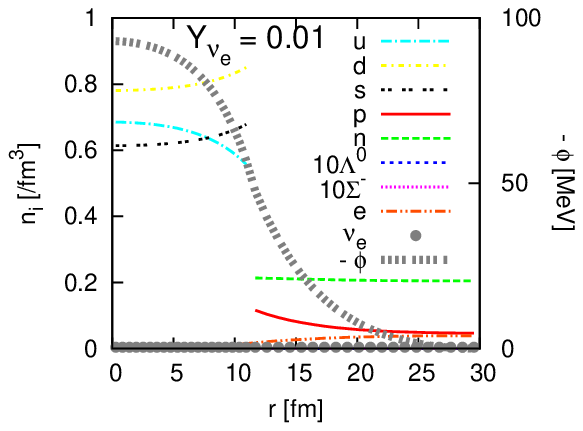}
\includegraphics[width=20pc]{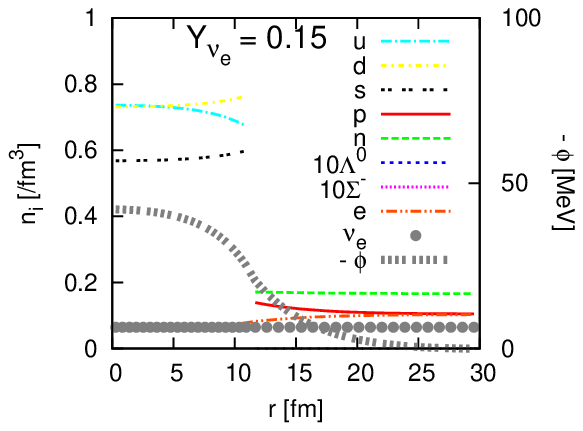}
\caption{\label{label03}Density profiles and Coulomb potential $\phi$ within 1D (slab) for $n_B = 2.5~n_0$ at $T=10$ MeV. Here, the neutrino fractions are set to  $Y_{\nu_e}=0.01$~(left panel) and $Y_{\nu_e}=0.15$~(right panel). The cell sizes are $R_W = 30$ fm in these figures. The slab size are $R = 10.9 $ fm. $(R/R_W)^3$ is fixed to be the optimal value at $Y_{\nu_e} = 0.01$.}
\end{figure}

Fig.~\ref{label03} shows the change of the Coulomb energy by trapped neutrinos more clearly. The figure shows the density profiles within the 1D cell (slab) for $n_B = 2.5~n_0$ at $Y_{\nu_e} = 0.01$~(left panel) and $Y_{\nu_e} = 0.01$~(right panel). Each temperature is set to be $T=10$ MeV. The volume ratio $(R/R_W)^3$ of the quark phase is fixed to be the optimal value at $Y_{\nu_e} = 0.01$.  Clearly, the magnitude of the Coulomb energy is decreased at high neutrino fraction. 
At high neutrino fraction, since the enhancement of electrons increases the particle fractions with positive charge, the negative charge of the quark slab is suppressed. As the result, the magnitude of the Coulomb energy by the quark slab is decreased. Of course, the particle fractions with positive charge increase in the hadron slab, too. But, they do not change so much because of their heavy masses. 

%----------------------
\section{Summary and Discussion}
%----------------------
We have studied the hadron-quark phase transition and the properties of the mixed phase in proto-neutron stars. We have taken into account the finite-size effects by imposing the Gibbs conditions on the phase 
equilibrium, and calculating the density profiles in a self-consistent manner.

We find that the hadron-quark mixed phase become unstable at high temperature and/or high lepton fraction. The instability by finite temperature mainly comes from the correlation energy, while the effects of neutrinos on free energy appear both in the correlation energy and the Coulomb energy.

Hence, we can suggest that the quark-phase transition is close to the EOS with the Maxwell condition at the very first stage of proto-neutron stars, because the temperature is high, $T \sim 40$ MeV, and the neutrino fractions are also high, $Y_{\nu_e} \sim 0.2$.
However, in the middle stage of evolution of proto-neutron stars, the pasta structure will appear, where initial cooling and deleptonization proceed. 
Hence, the simple Maxwell condition might be enough to take into account the quark-hadron phase transitions in the simulations of supernovae or black hole formations, although such treatment is inappropriate in the simulations of proto-neutron star evolutions.
We can see the similar discussions in the recent papers by Schaffner's Group, though they have not included the finite-size effects~\cite{pagliara09, hampel09}. 

Finally, we note again that the EOS has many uncertainties, especially for quark matter. We simply adopted the MIT bag model and the density-dependent bag model in this paper,  while other quark models, including chiral restoration or color superconductivities, may change our results~\cite{yasutake09a}. These are open questions for astrophysics and nuclear physics.

%------------------------
\section*{Acknowledgement}
%------------------------
We are grateful to S. Chiba, H.-J. Schulze, and M.~Baldo for their warm hospitality and fruitful discussions. This work was partially supported by the Grant-in-Aid for the Global COE Program ``The Next Generation of Physics, Spun from Universality	and Emergence,'' from the Ministry of Education, Culture, Sports, Science and Technology (MEXT) of Japan, and the Grant-in-Aid for Scientific Research (C) (20540267, 21105512, 19540313).

%------------------------
\section*{References}
%------------------------
%\bibliography{iopart-num}

\begin{thebibliography}{99}


\bibitem{glendenning92}
Glendenning N~K 1992 {\em Phys. Rev.\/} D {\bf 46} 1274

\bibitem{voskresensky02}
Voskresensky D~N, Yasuhira M and Tatsumi T 2002 {\em Phys. Lett.\/} B {\bf 541}
  93

\bibitem{endo06}
Endo T, Maruyama T, Chiba S and Tatsumi T 2006 {\em Prog. Theor. Phys.\/} {\bf
  115} 337

\bibitem{maruyama07}
Maruyama T, Chiba S, Schulze H-J and Tatsumi T 2007 {\em Phys. Rev.\/} D {\bf
  76} 123015

\bibitem{maruyama08}
Maruyama T, Chiba S, Schulze H-J and Tatsumi T 2008 {\em Phys. Lett.\/} B {\bf 
  659} 192

\bibitem{yasutake09b}
Yasutake N, Maruyama T and Tatsumi T 2009 {\em Phys. Rev.\/} D {\bf 80} 123009

\bibitem{baldo98}
Baldo M, Burgio G~F and Schulze H-J 1998 {\em Phys. Rev.\/} C {\bf 58} 3688

\bibitem{noumi02}
Noumi H, Saha P~K, Abe D, Ajimura S, Aoki K, Bhang H~C, Endo T, Fujii Y, Fukuda
  T, Guo H~C, Imai K, Hashimoto O, Hotchi H, Kim E~H, Kim J~H, Kishimoto T,
  Krutenkova A, Maeda K, Nagae T, Nakamura M, Outa H, Sekimoto M, Saito T,
  Sakaguchi A, Sato Y, Sawafta R, Shimizu Y, Takahashi T, Tang L, Tamura H,
  Tanida K, Watanabe T, Xia H~H, Zhou S~H, Zhu L~H and Zhu X~F 2002 {\em Phys.
  Rev. Lett.\/} {\bf 89} 072301

\bibitem{saha04}
Saha P~K, Noumi H, Abe D, Ajimura S, Aoki K, Bhang H~C, Dobashi K, Endo T,
  Fujii Y, Fukuda T, Guo H~C and Hashimoto O 2004 {\em Phys. Rev.\/} C {\bf 70}
  044613

\bibitem{baldo99}
Baldo M and Ferreira L~S 1999 {\em Phys. Rev.\/} C {\bf 59} 682

\bibitem{nicotra06b}
{Nicotra} O~E, {Baldo} M, {Burgio} G~F and {Schulze} H-J 2006 {\em Phys.
  Rev.\/} D {\bf 74} 123001

\bibitem{yasutake09a}
Yasutake N and Kashiwa K 2009 {\em Phys. Rev.\/} D {\bf 79} 043012

\bibitem{burrows86}
Burrows A and Lattimer J~M 1986 {\em Astro. Phys. Jour.\/} {\bf 307} 178

\bibitem{pagliara09}
Pagliara G, Hempel M and Schaffner-Bielich J 2009 {\em Phys. Rev. Lett.\/} {\bf
  103} 171102

\bibitem{hampel09}
Hempel M, Pagliara G and Schaffner-Bielich J 2009 {\em Phys. Rev.\/} D {\bf 80}
  125014
\end{thebibliography}

\end{document}